\documentclass[twocolumn,nofootinbib,amsmath,amssymb,a4paper]{revtex4}

\RequirePackage{xspace}
\usepackage{graphicx}
\usepackage{dcolumn}
\usepackage{bm}
\usepackage{relsize}
\def\babar{\mbox{\slshape B\kern-0.1em{\smaller A}\kern-0.1em
    B\kern-0.1em{\smaller A\kern-0.2em R}}}
\def\Kbar{\kern 0.2em\overline{\kern -0.2em K}{}\xspace}
\def\Kzb{\Kbar^0\xspace}
\def\Bbar    {\kern 0.18em\overline{\kern -0.18em B}{}\xspace}
\def\Bzb     {\ensuremath{\Bbar^0}\xspace}

\begin{document}

\title{$B^0\to K^0 \Kzb$ and Other Hadronic $b \to d$ Decays}

\author{J. Biesiada}
 \email{biesiada@princeton.edu}
\affiliation{%
Department of Physics, Princeton University, Princeton, NJ  08544, USA\\
}%

\begin{abstract}
The $b \to d$ penguin-dominated modes $B \to K \overline{K}$ have been observed at the $B$ factories.  In addition, the \babar\ collaboration has reported the first time-dependent $CP$-violation measurement in $B^0\to K^0 \Kzb$.
\end{abstract}

\maketitle

\section{Introduction}
These proceedings summarize measurements of $B$ decays to two kaons at the $B$ factories~\cite{bib:BaBarResults}~\cite{bib:BelleResults}.  
Observations of the $B^0 \to K^0 \Kzb$ and $B^+ \to \Kzb K^+$ decays 
have been reported by the \babar\ experiment~\cite{bib:babar} at
 SLAC and the Belle experiment~\cite{bib:belle} at KEK.  The \babar\ collaboration has 
also performed the first time-dependent $CP$-violation measurement in $B^0\to K^0 \Kzb$.
  The $B^+ \to K^+ K^-$ decay has not been observed by either experiment and the upper limits on its branching fraction were improved.

\begin{figure}
\includegraphics[width=0.27\textwidth]{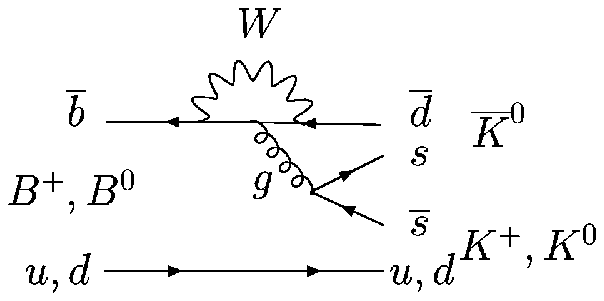}
\includegraphics[width=0.2\textwidth]{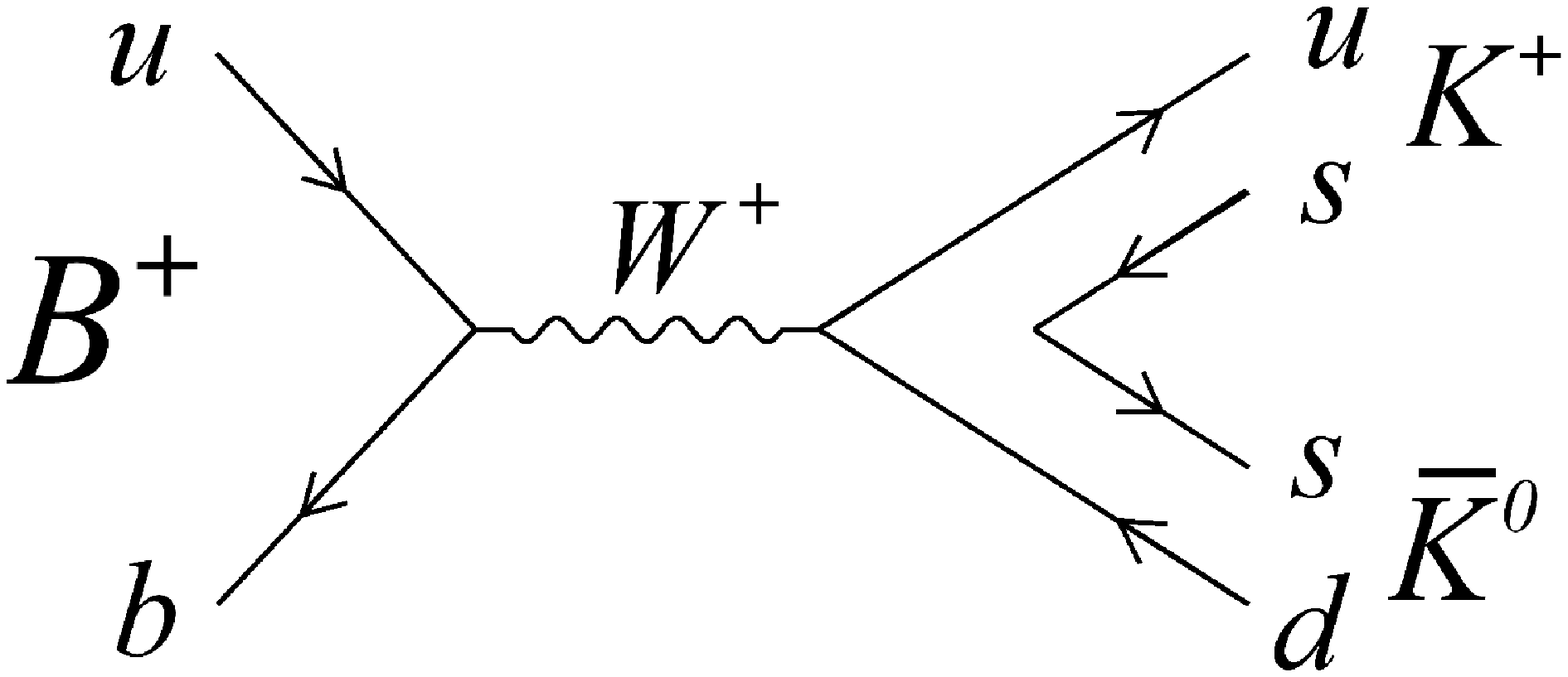}
\caption{\label{fig:btodpenguin} The $b \to d \bar{s} s$ penguin transition (left) and the
annihilation amplitude in $B^+ \to \Kzb K^+$ (right).}
\end{figure}

$B \to K \overline{K}$ decays are expected to be dominated by the $b \to d \bar{s} s$ amplitude
 involving a virtual loop with the emission of a gluon (Fig.~\ref{fig:btodpenguin}).  These decays are suppressed 
in the Standard Model (SM), with branching fractions at the $10^{-6}$ level.  
Assuming top-quark dominance in the virtual loop~\cite{bib:Fleischer94}, the direct and mixing-induced 
$CP$ asymmetries are expected to vanish in these modes.  However, contributions from 
up and charm quarks as well as from non-SM particles could induce sizeable CP asymmetries~\cite{bib:Giri}.  
These modes are thus analogous to the $b \to s \bar{s} s$ decays in the measurement of virtual 
effects of contributions from non-SM phenomena~\cite{bib:LondonQuinn,bib:phiK}.
  Additional amplitudes could affect 
$b\to d$ penguin decays differently from $b\to s$ penguin decays. In addition, 
various relations between branching fractions and CP asymmetries in these modes provide 
a test of SM predictions~\cite{bib:FleischerRecksiegel}.  Finally, the $B^+ \to \Kzb K^+$ decay has an 
annihilation contribution in the SM (Fig.~\ref{fig:btodpenguin}).  Thus, the difference in decay rate between the charged
 mode and the neutral mode can be used to constrain the effect of this amplitude in $B$ decays.

\section{Experimental Methods}

\subsection{Signal Extraction}
Signal decays are separated from background decays using unbinned
extended maximum likelihood fitting to distributions of kinematic 
and event-shape variables.  The primary kinematic variables used to identify
a reconstructed signal $B$ candidate are the difference between its reconstructed 
energy in the center-of-mass (CM) frame and the beam energy; and a 
beam-energy substituted mass, $m_{\rm ES} \equiv\sqrt{(s/2+{\mathbf {p}}_i
\cdot{\mathbf {p}}_B)^2/E^2_i-{\mathbf {p}}^2_B}$, where the $B$-candidate momentum
 ${\mathbf {p}}_B$ and the four-momentum of the initial $e^+e^-$ state 
$(E_i, {\mathbf {p}}_i)$ are calculated in the laboratory frame.  Event-shape variables
are used to suppress the dominant ``continuum'' $e^+e^- \to q\bar{q}$ $(q=u,d,s,c)$ 
background further, exploiting angular differences between 
the jet-like topology of continuum decays and the isotropically distributed decays 
of $B \overline{B}$ events.  Particle-identification information is used to separate
charged pion from charged kaon candidates in the $B^+ \to \Kzb K^+$ and 
$B^+ \to K^+ K^-$ decays.

\subsection{Time-Dependent CP Asymmetries}
$CP$ asymmetries in $B^0 \to K^0 \Kzb$ are determined from the difference
in the time-dependent decay rates $f_+$ and $f_-$ for $\Bzb$ and $B^0$ signal decays, respectively,
where

\begin{equation}
 f_{\pm}(\Delta t) = \frac{e^{-\left|\Delta t\right|/\tau}}{4\tau} [1
\pm S\sin(\Delta m_d \Delta t) \nonumber \\
\mp C\cos(\Delta m_d \Delta t)].
\label{eq:eq1}
\end{equation}
\\
Here $\Delta t$ is the time difference between the decays of
the signal $B$ and the other $B$ in the event, $\tau$ is the average $B^0$ lifetime,
and $\Delta m_d$ is the $B^0-\Bzb$ mixing frequency.
The amplitude $S$ describes $CP$ violation in the interference
between mixed and unmixed decays into the same final state, while $C$ describes
direct $CP$ violation in decay.  The time-dependent distribution is 
corrected for detector resolution effects and for mistag rates
 in the flavor identification of the other $B$ in the event.

$B^0 \to K^0 \Kzb$ decays are reconstructed in the $K^0 \Kzb \to K_S K_S$
 and $K_S \to \pi^+\pi^-$ sub-decay modes.  As the $K_S$ meson has a relatively long lifetime of $90$ ps,
its decay vertex is separated from the $B$ decay vertex by a few centimeters, which complicates
the measurement of the $z$ position of the signal $B$ in the absence of prompt charged tracks 
originating from its decay vertex.  To perform the measurement,
\babar\ used a method previously employed in
other $K_S$ analyses without prompt charged tracks, 
wherein the $B$ meson is constrained in the vertex fit
to decay within the beamspot in the plane transverse to the beam direction~\cite{bib:KsVertexing}.
This method exploits the threshold nature of the $\Upsilon (4S) \to B\overline{B}$ decay,
 where the $B$ mesons are almost at rest in the CM frame (${\mathbf p^{\rm *}_{\rm B}}\approx 300$ MeV/c)
and therefore have negligible displacement in the lab frame in the non-boosted transverse direction.
The resulting resolution of the $z$ position of the signal $B$ vertex is still better than the resolution on
the vertex position of the other $B$ in the event, yielding a $\Delta t$ precision (approximately $0.9$ ps) comparable
to that in modes where the signal $B$ has prompt charged tracks in the final state.  Only $K_S$ mesons
that decay within the Silicon Vertex Tracker can be used with this technique.  However, as there are two 
$K_S$ mesons in the final state, while only one is needed to employ this method, the fraction of signal decays
suitable for the time-dependent measurement at \babar\ is approximately $82\%$ in this mode.

If no time-dependent measurement is performed, an integrated flavor or charge asymmetry can be measured:

\begin{equation}
{\cal A}_{CP} = \left(N_{B^0,B^+} - N_{\Bzb,B^-}\right)/\left(N_{B^0,B^+} + N_{\Bzb,B^-}\right)
\end{equation}
\\
A non-zero value of this asymmetry signifies the presence of direct $CP$ violation.
In the $B^+$ modes, this is the only possible $CP$ measurement.

\section{Experimental Results}

\subsection{$B^0 \to K^0 \Kzb$}

\begin{figure*}
\includegraphics[width=0.5 \textwidth]{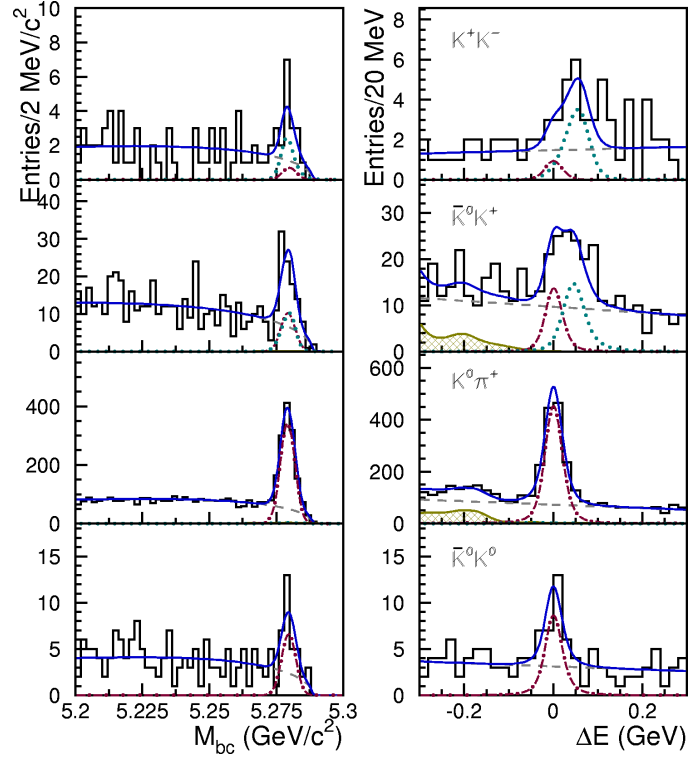}
\caption{Distributions of $m_{\rm ES}$ (left) and $\Delta E$ (right) in data (solid histogram) from the Belle experiment with the results of the fit overlaid for signal (dot-dashed), continuum (dashed), charmless B decays (hatched), feed-across background from misidentification (dotted), and the sum of all components (solid).}
\label{fig:BellePlot}
\end{figure*}

\begin{figure*}
\includegraphics[width=0.4 \textwidth]{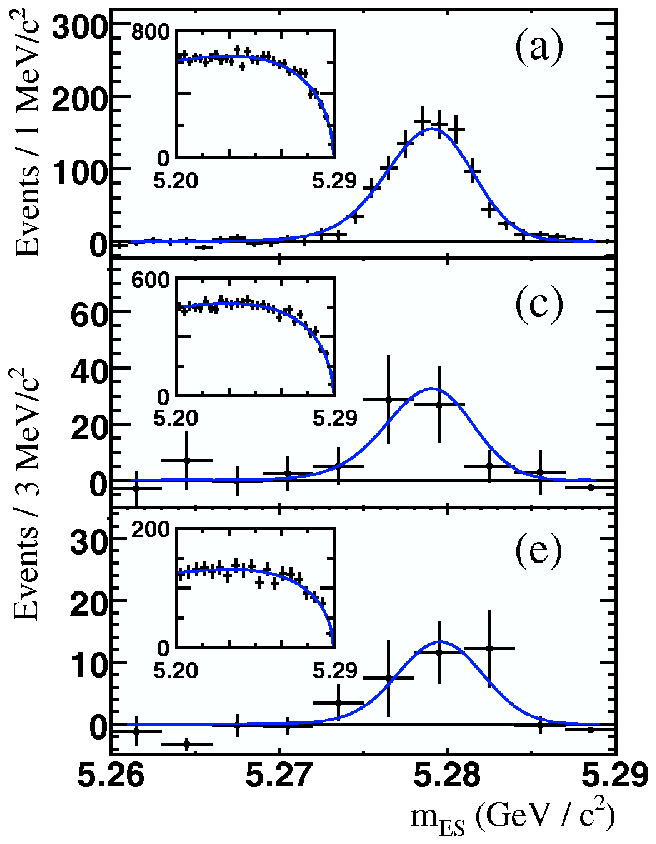}
\includegraphics[width=0.4 \textwidth]{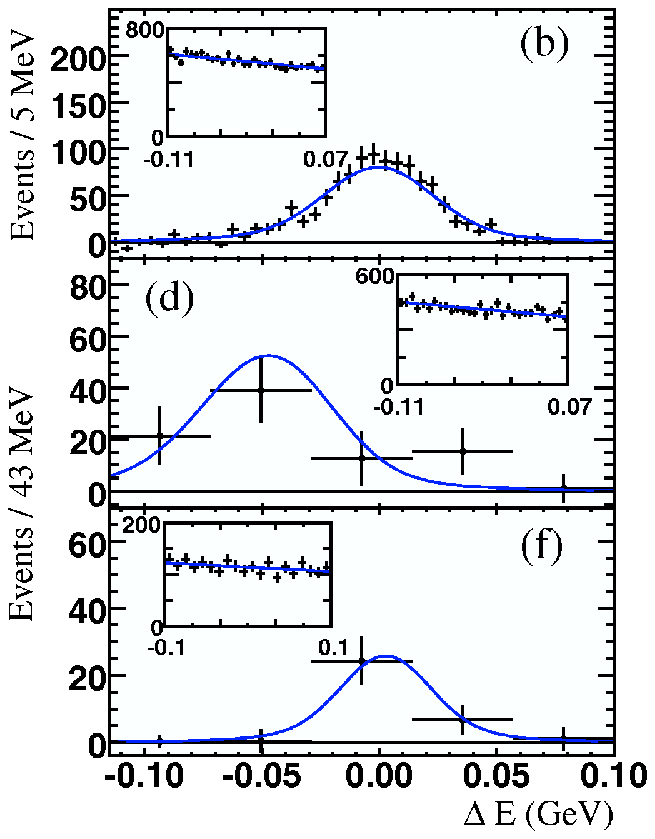}
\caption{Distributions of $m_{\rm ES}$ (left) and $\Delta E$ (right) 
in data (points with error bars) from the \babar\ experiment for signal (main plots) 
and background (insets) for $K^0 \pi^+$ (a,b), $\Kzb K^+$ (c,d), 
and $K^0 \Kzb$ (e,f) candidates, with the results of the fit overlaid (solid lines).  
The distributions are obtained with the sPlot technique~\cite{bib:sPlots}.}
\label{fig:BabarPlot}
\end{figure*}

Both $B$ factories observe clear signals in this decay mode,
 with a branching fraction at the $10^{-6}$ level, in agreement
 with each other and Standard-Model expectations.   \babar\ observes
 the signal at the $7.3\sigma$ level of significance, 
including systematic uncertainties, while Belle's observation is at
 the $5.3\sigma$ level.  The branching fraction results are summarized
 in Table~\ref{tab:KsKs}, while projections of the fitted data in the $m_{\rm ES}$ and $\Delta E$ variables with the results of the fit overlaid are shown in Figs.~\ref{fig:BellePlot} and~\ref{fig:BabarPlot}.

\begin{figure}
\includegraphics[width=0.48 \textwidth]{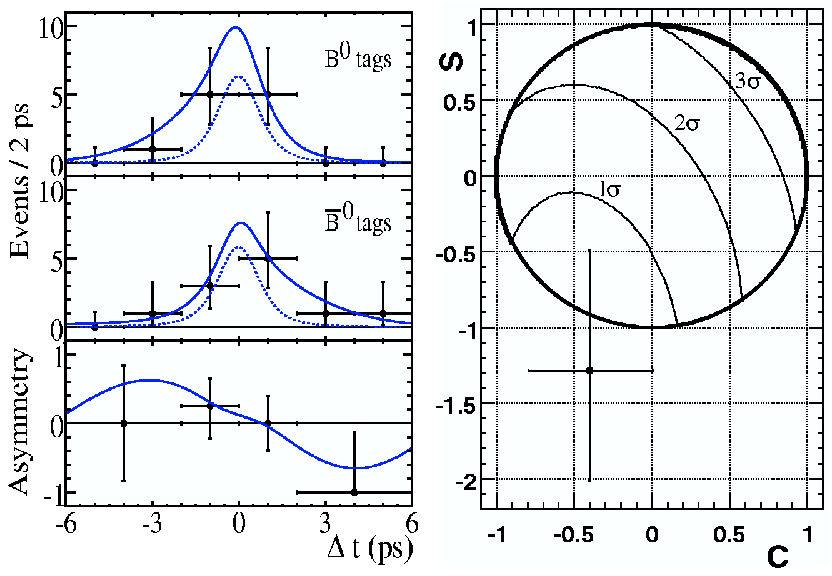}
\caption{The left plots show the distributions of $\Delta t$ in $\babar~{\rm 's}$ time-dependent fit to data for decays tagged as $B^0$ (top) and $\Bzb$ (middle), as well as the asymmetry (bottom).  The dotted line shows the PDF projection of the fit for the background component, while the solid line is the sum of signal and background.  The plot is enhanced in signal decays using additional selections on probability ratios of the variables not shown in the plot.  The right plot shows the result (point with error bars) and the physically allowed region (solid circle) in the $S$ versus $C$ plane.  The likelihood contours correspond to changes in $-2\log{\cal L}$
of $2.3$ for $n=1$,
$6.2$ for $n=2$, and $11.8$ for $n=3$.}
\label{fig:BabarCP}
\end{figure}

The results of \babar\'s time-dependent $CP$ measurement are presented
 in Table~\ref{tab:KsKs} and Fig~\ref{fig:BabarCP}.  While the errors are still large,
 the measurement already excludes a fraction of the physically allowed region
 at a confidence level of greater than $3\sigma$, as seen in 
Fig.~\ref{fig:BabarCP}. In particular, large positive values of $S$ are disfavored,
although more data is needed to confirm this result.
Belle did not perform a time-dependent measurement, 
but measured the time-integrated CP asymmetry, 
reporting ${\cal A}_{CP}\footnote{${\cal A}_{CP}=-C$}=-0.58^{+0.73}_{-0.66}\pm0.01$.

\begin{table}
\caption{
Experimental results on $B^0 \to K^0 \Kzb$ decays, with the number of $B\overline{B}$ pairs
listed in the last column.
}
\label{tab:KsKs}
\begin{tabular}{l|lc}
\hline
Experiment & Observable & $N_{B\overline{B}}\times 10^{6}$ \\
\hline
\babar\   & ${\cal B}= (1.08\pm0.28\pm0.11)\times 10^{-6}$ & $347$ \\
        & $S= -1.28^{+0.80}_{-0.73}\,^{+0.11}_{-0.16}$ \\
        & $C= -0.40 \pm0.41\pm0.06$ \\
\hline
Belle   & ${\cal B}= (0.87 ^{+0.25}_{-0.20} \pm 0.09)\times 10^{-6}$ & $449$ \\
        & ${\cal A}_{CP}= -0.58^{+0.73}_{-0.66}\pm0.04 $ \\
\hline
\end{tabular}
\end{table}

\subsection{$B^+ \to \Kzb K^+$ and $B^+ \to K^0 \pi^+$}

Both $B$ factories observe clear signals in this decay mode, with
 a branching fraction at the $10^{-6}$ level and a significance of $5.3\sigma$
in each experiment.  The branching fractions are slightly higher than the corresponding 
values for the neutral mode in each experiment.  Although the errors
 are too large at this point to make a substantive comparison, if this 
discrepancy holds up it would indicate a significant effect from 
the annihilation amplitude in the charged mode.  The results are 
summarized in Table~\ref{tab:KsK}, while projections of the fitted data
 in the $m_{\rm ES}$ and $\Delta E$ variables with the results of the fit
 overlaid are shown in Figs.~\ref{fig:BellePlot} and~\ref{fig:BabarPlot}.
  The $B^+ \to K^0 \pi^+$ mode has the same
topology as the $B^+ \to \Kzb K^+$ mode, and its branching fraction is
determined from the same fit, separating the two modes with particle ID.
The results are in Table~\ref{tab:KsPi}. Neither experiment observes
 a significant direct $CP$ asymmetry in either mode.

\begin{table}
\caption{
Experimental results on $B^+ \to \Kzb K^+$ decays, with the number of $B\overline{B}$ pairs
listed in the last column.
}
\label{tab:KsK}
\begin{tabular}{l|lc}
\hline
Experiment & Observable & $N_{B\overline{B}}\times 10^{6}$ \\
\hline
\babar\   & ${\cal B}= (1.61\pm0.44\pm0.09)\times 10^{-6}$ & $347$ \\
        & ${\cal A}_{CP}= 0.10\pm0.26\pm0.03$ \\
\hline
Belle   & ${\cal B}= (1.22^{+0.32}_{-0.28}\,^{+0.13}_{-0.16})\times 10^{-6}$ & $449$ \\
        & ${\cal A}_{CP}= 0.13^{+0.23}_{-0.24}\pm0.02 $ \\
\hline
\end{tabular}
\end{table}

\subsection{$B^0 \to K^+ K^-$}

This mode is dominated by a $W$-exchange amplitude and is not expected to be
seen at the $B$ factories, although long-distance rescattering and effects from
beyond-SM physics could affect this conclusion.  The expected SM branching fractions
are expected at the $10^{-8}$ to $10^{-7}$ level.  Neither experiment sees
any evidence for this mode, and upper limits are set as summarized in 
Table~\ref{tab:KK}~\cite{bib:BelleResults}~\cite{bib:BaBarKK}.

\begin{table}
\caption{
Experimental results on $B^+ \to K^0 \pi^+$ decays, with the number of $B\overline{B}$ pairs
listed in the last column.
}
\label{tab:KsPi}
\begin{tabular}{l|lc}
\hline
Experiment & Observable & $N_{B\overline{B}}\times 10^{6}$ \\
\hline
\babar\   & ${\cal B}= (23.9\pm1.1\pm1.0)\times 10^{-6}$ & $347$ \\
        & ${\cal A}_{CP}= -0.029\pm0.039\pm0.010$ \\
\hline
Belle   & ${\cal B}= (22.8^{+0.8}_{-0.7} \pm 1.3)\times 10^{-6}$ & $449$ \\
        & ${\cal A}_{CP}= 0.03\pm 0.03 \pm 0.01 $ \\
\hline
\end{tabular}
\end{table}

\begin{table}[!h]
\caption{
Experimental results on $B^0 \to K^+ K^-$ decays, with the number of $B\overline{B}$ pairs
listed in the last column.
}
\label{tab:KK}
\begin{tabular}{l|lc}
\hline
Experiment & Observable & $N_{B\overline{B}}\times 10^{6}$ \\
\hline
\babar\   & ${\cal B}< 0.5 \times 10^{-6}$ {\text at} $90\% ~C.L.$ & $227$ \\
\hline
Belle   &  ${\cal B}< 0.41 \times 10^{-6}$ {\text at} $90\% ~C.L.$ & $449$ \\
\hline
\end{tabular}
\end{table}

\section{Future Outlook}

Each of the $B$ factories is expected to accumulate $1~{\rm ab}^{-1}$ by the end of operations in $2008$.
  With these datasets, the error on the branching fractions of $B^0 \to K^0 \Kzb$ and
 $B^0 \to \Kzb K^+$ should decrease to the $15\%$ level for each experiment.  
Also, each collaboration should achieve an error of $\sim$$0.4$ on $S$ and $\sim$$0.25$ on $C$
 in $B^0 \to K^0 \Kzb$.
  These results will provide meaningful comparisona with theoretical predictions and 
constrain contributions from beyond-SM physics in $b \to d$ FCNC decays.  
The effect of the annihilation contribution in $B^0 \to \Kzb K^+$ 
will be constrained as well.

\section{Summary}

The $B$ factories have observed the $b \to d$ penguin dominated modes 
$B^0 \to K^0 \Kzb$ and $B^0 \to \Kzb K^+$.  The measured branching fractions are 
consistent with Standard Model predictions, although experimental errors are 
still too large for precision comparisons.  \babar\ has also performed the first
 time-dependent $CP$ violation measurement in $B^0 \to K^0 \Kzb$, 
excluding a part of the physically allowed region
 at greater than $3\sigma$ significance.  This is the first step in the investigation of 
$CP$ violation in $b \to d$ flavor-changing neutral-current decays.


\begin{thebibliography}{99}

\bibitem{bib:BaBarResults}
\babar\ Collaboration, B.~Aubert {\em et al.}, {\em Phys. Rev. Lett.} 97, 171805 (2006).

\bibitem{bib:BelleResults}
Belle Collaboration, hep-ex/0608049.

\bibitem{bib:babar}
\babar\ Collaboration, B.~Aubert {\em et al.}, {\em Nucl. Instrum. Meth. A} 
{\bf 479}, 1 (2002).

\bibitem{bib:belle}
Belle Collaboration, A.~Abashian {\em et al.}, {\em Nucl. Instrum. Meth. A} 
{\bf 479}, 117 (2002).

\bibitem{bib:Fleischer94}
R.~Fleischer, {\em Phys. Lett. B} 341, 205 (1994).

\bibitem{bib:Giri}
A.~K.~Giri and R.~Mohanta, {\em J. of High Energy Phys.} {\bf 11}, 084 (2004).

\bibitem{bib:LondonQuinn}
D.~London and R.~D.~Peccei, {\em Phys. Lett. B} 223, 257 (1989); 
H.~R.~Quinn, {\em Nucl. Phys. B, Proc. Suppl.} 37A, 21 (1994).

\bibitem{bib:phiK}
\babar\ Collaboration, B.~Aubert {\em et al.},  {\em Phys. Rev. Lett.} 71, 091102 (2005); 
Belle Collaboration, K.~Abe {\em et al.}, {\em Phys. Rev. Lett.} 91, 261602 (2003).

\bibitem{bib:FleischerRecksiegel}
R. Fleischer and S. Recksiegel, {\em Eur. Phys. J. C} 38, 251 (2004).

\bibitem{bib:KsVertexing}
\babar\ Collaboration, B.~Aubert {\em et al.}, {\em Phys. Rev. D} 71, 111102 (2005).

\bibitem{bib:sPlots}
M.~Pivk and F.~R.~Le~Diberder, {\em Nucl. Instrum. Meth. A} 555, 356 (2005).

\bibitem{bib:BaBarKK}
\babar\ Collaboration, B.~Aubert {\em et al.}, {\em Phys. Rev. D} 75, 012008 (2007).

\end{thebibliography}
\end{document}